\documentclass[aps,prl,twocolumn,superscriptaddress]{revtex4-1}

\usepackage{amsmath,amssymb,bm}
\usepackage{graphicx}
\usepackage{epstopdf}
\usepackage{color} 
\DeclareMathOperator{\sinc}{sinc}
\begin{document}


\newcommand{\todo}[1]{\textcolor{red}{\textit{{#1}}}}
\newcommand{\Four}[1]{\mathcal{F}\left[{#1}\right]}
\newcommand{\MagFour}[1]{\left|\mathcal{F}\left[{#1}\right]\right|^2}
\newcommand{\FourInv}[1]{\mathcal{F}^{-1}\left[{#1}\right]}
\newcommand{\FourNoArg}{\mathcal{F}}
\newcommand{\Expect}[2]{\mathbb{E}_{p({#1})}\left[{#2}\right]}
\newcommand{\rect}{\textnormal{rect}}
\renewcommand{\Re}{\textnormal{Re}}

\title{Narrowband oscillations from asynchronous neural activity}


\author{Stephen V. Gliske}
\email[]{sgliske@umich.edu}
\affiliation{Dept. of Neurology, University of Michigan}
\author{Eugene Lim}
\affiliation{Dept. of Physics, Ohio Wesleyan University}
\author{Katherine A. Holman}
\affiliation{Dept. of Physics, Towson University}
\author{William C. Stacey}
\affiliation{Dept. of Neurology, University of Michigan}
\affiliation{Dept. of Biomedical Engineering, University of Michigan}
\author{Christian G. Fink}
\affiliation{Dept. of Physics, Ohio Wesleyan University}
\affiliation{Neuroscience Program, Ohio Wesleyan University}


\date{\today}

\begin{abstract}
We investigate the possibility that narrowband
oscillations may emerge from completely asynchronous, independent
neural firing. We find that a population of asynchronous neurons may
produce narrowband oscillations if each neuron fires
quasi-periodically, and we deduce bounds on the degree of variability
in neural spike-timing which will permit the emergence of such
oscillations. These results suggest a novel mechanism of neural
rhythmogenesis, and they help to explain recent experimental reports
of large-amplitude local field potential oscillations in the absence
of neural spike-timing synchrony. Simply put,
  although synchrony can produce oscillations, oscillations do not
  always imply the existence of synchrony. 
\end{abstract}

\pacs{}

\maketitle


Neural rhythms, as observed in electroencephologram (EEG) and local field potential (LFP) recordings, are associated with various brain functions and are generated through manifold mechanisms \cite{buzsaki2004neuronal}. One interesting feature of neural rhythms is that they are often observed in conjunction with irregular spiking of individual neurons \cite{wang2010neurophysiological}. This phenomenon has previously been explained by analyzing the interplay between excitatory and inhibitory synaptic time scales and feedback loops \cite{brunel2003determines}, stochastic resonance \cite{moss2004stochastic}, or correlations in stochastic input \cite{doiron2004oscillatory}. However, all of these mechanisms will produce non-trivial levels of spike-timing synchrony within a population of neurons, and recent experimental studies have demonstrated examples of epileptic seizures which feature narrowband LFP oscillations in the absence of population synchrony \cite{alvarado2013single,truccolo2014neuronal}. 

In the present work we demonstrate that narrowband oscillations may emerge from a population of neurons that fire asynchronously, independently, and stochastically. This may be accomplished if the neurons naturally fire with some rhythmicity and with similar average frequencies, conditions which may plausibly be met if a population of neurons receives similar intensity of input and shares similar biophysical parameters. This work therefore proposes a novel and general mechanism for the generation of brain rhythms. We also derive bounds on the levels of spike-timing heterogeneity which allow for the emergence of such rhythms.

As a toy example, consider a situation in which a population of $N$ neurons fire with the same frequency $f_0$, but with uniformly random phase. The contribution to the LFP by any one neuron, $g(t)$, is well approximated as the convolution of a periodic train of delta functions with a kernel waveform (representing the voltage trace of an individual action potential, for example). The Fourier transform of this signal, $G(f)$, will feature peaks at $f_0$ and its harmonics, with an amplitude of zero at all other frequencies. The LFP will then be the superposition of $N$ randomly-shifted versions of $g(t)$,
$g_N(t) = \sum^N_{j=1} g(t-t_{0,j})$, with  $t_{0,j} \sim  \mathrm{unif}(0,\frac{1}{f_0})$. The Fourier transform of the LFP is then $G_N(f) = \left[ \sum^N_{j=1} e^{i \theta_j} \right] G(f)$, where $\theta_j = -2 \pi f t_{0,j}$. The energy spectral density can be determined by defining $A = \sum^N_{j=1} e^{i \theta_j}$ and computing its expected squared amplitude:

\begin{align}
  E\left\{ |A(f)|^2 \right\} 
    =& \int_0^{1/f_0} dt_{0,1} \dots dt_{0,N}\ p(t_{0,1})\;\dots\; p(t_{0,N})\ |A(f)|^2 \nonumber
\end{align}
\begin{align}
    &= \left(\frac{-f_0}{2\pi f}\right)^N \int_0^{2\pi f/f_0} d\theta_1 \dots d\theta_N \left( \sum_{j=1}^N e^{i \theta_j} \right) \left( \sum_{k=1}^N e^{-i \theta_k} \right) \nonumber\\
    &= N + N(N-1) \sinc^2\left(\frac{\pi f}{f_0}\right) \nonumber
\end{align}
Combining this result with the fact that $G_N(f)=0$ for all non-harmonic frequencies, the energy spectral density is
\begin{equation}
E \left\{|G_N(f)|^2 \right\} = \begin{cases}
	N  |G(f)|^2, &  \text{$f=n f_0$ ($n=1, 2, 3...$)} \\
	0, & \text{otherwise}
\end{cases}
\label{eq:final_eq}
\end{equation}

This toy model therefore suggests the possibility of narrowband collective oscillations emerging from asynchronous neural activity. This example is analogous to the fact that incoherent light waves do not produce completely destructive interference, but superimpose with an intensity that scales linearly with the number of waves.

Of course, individual neurons do not spike perfectly periodically, nor do they share the same intrinsic frequency across a population. We therefore introduce a model of asynchronous, independent, and stochastic neural activity which takes both of these sources of spike-time variability into account. Specifically, we consider a superposition of renewal processes (which is not itself a renewal process \cite{lindner2006superposition}) in which the inter-event interval (IEI) density of neuron $i$ is given by 
\vspace{-5mm}
\begin{equation}
p_0(\tau^{(i)}|\mu_i, \sigma_{\rm{jit}})=\frac{1}{\sigma_{\rm{jit}}\sqrt{2 \pi} } \; \mathrm{exp}\left( \frac{-(\tau^{(i)}-\mu_i)^2}{2 \sigma_{\rm{jit}}^{2}} \right),
\label{eq:iei}
\end{equation}
\noindent
with the parameter $\mu_i$ also being normally distributed, drawn from
$\mathcal{N}( \mu_0, \sigma_\mu)$, and
  $\sigma_{\rm{jit}}$ being a fixed model parameter.  Therefore
$\sigma_\mu$ determines the variability in intrinsic frequency for the entire
population, and $\sigma_{\rm{jit}}$ quantifies the degree of
``jitter'' from one event to the next for a single neuron. The mean
population frequency is set by $\mu_0$. (Note that while this
technically permits negative IEI values, this will occur very rarely
as long as $\sigma_\mu$ and
$\sigma_{\rm{jit}}$ are kept sufficiently small with respect to $\mu_0$.)

We assume all events generate either an action potential (AP) or post-synaptic potential (PSP) voltage waveform, 
so that the overall LFP is computed as the convolution of the
waveform with the event trains generated by Eq. \ref{eq:iei}, summed over all neurons. Our goal is to compute
the expected energy spectral density of this model LFP. Since the energy spectrum of such
a convolution is the product of the energy spectra of the fixed
waveform and the event train, we initially focus on just the spectrum 
of the event train.

Let each neuron have event times given by
$f^{(k)}(t)$, with the population level spike train being
  $f_T(t) = \sum_{j=1}^{N_C} f^{(j)}(t)$
and $N_C$ being the number of cells.
The energy spectral density of this aggregate event train is then given by the functional integral
\vspace{-3mm}
\begin{eqnarray}
  \Expect{f}{\left|\Four{f_T}\right|^2}
     &=& \int df_T\ p(f_T) \left|\Four{f_T}\right|^2,\nonumber \\ 
     &=& \int \left[ \prod_j df^{(j)}\ p\left(f^{(j)}\right)\right] \left|\sum_{k=1}^{N_C} \Four{f^{(k)}}\right|^2. \nonumber \\ \nonumber
\end{eqnarray}
Assuming event trains are independent from cell to cell, and that all cells' event trains are drawn
from the same family of distributions, this simplifies to
\begin{eqnarray}
  \Expect{f}{\left|\Four{f_T}\right|^2}
  \label{eqn:func_integral_pop_3b}
     &=& N_C \int df\ p\left(f\right) \left| \Four{f}\right|^2
\\ && {} + 
          \frac{1}{2}N_C(N_C-1) \left| \int df\ p\left(f\right) \Four{f} \right|^2.
\nonumber
\end{eqnarray}


In general, since each event train is parameterized by a discrete set of event times $t_k$, the integration measures are 
  $\int df\ p(f) = \int d\bm\theta p(\bm\theta) \prod_k dt_k \ p_k( t_k | \bm\theta )$,
where $\bm\theta$ represents any additional model parameters on which
the pdfs are conditioned.
The Fourier transform of $f$
and its magnitude are then
\begin{eqnarray}
\Four{f} &=& \frac{1}{\sqrt{2\pi}} \frac{1}{N_S}  \sum_k e^{-i\omega t_k},\\
\MagFour{f} &=& \frac{1}{2\pi} \frac{1}{N_S^2}  \sum_{k,\ell} e^{-i\omega (t_k-t_\ell)}.
\end{eqnarray}

Applying the change of variables, $\tau_k = t_k - t_{k-1}$,
and making the standard renewal process assumption of IEI independence 
enables the application of an IEI density function, such as that defined in Eq. \ref{eq:iei}. Recalling the assumption that
events occur independently from cell to cell, we may formally state that
 $p_k^{(i)}\left(\left.\tau_k^{(i)}\right|\bm\theta_k^{(i)}\right) = p_0\left(\left.\tau^{(i)}\right| \bm\theta^{(i)} \right)$, 
for the $k$th event on the $i$th neuron.

In order to model asynchronous neural activity from the outset, we introduce a randomly distributed initial temporal offset, $t_0^{(i)}$, as one of the model parameters included in $\bm\theta^{(i)}$. Our model also assumes that inter-event intervals are centered around some value $\mu$, unique for each neuron, so that $\Four{p_0}$ can best be expressed as
   $\Four{p_0} = e^{-i \mu \omega} \Four{p_0^\prime}$.
The quantity $\bm\theta$ is thus the $n$-tuple $[t_0, \mu]$, with $p(\bm \theta) = p_{t_0}(t_0)\ p_\mu(\mu)$.

The above assumptions imply that in general 
\begin{eqnarray}
\int df \ p\left(f\right) \Four{f}
    &=& \frac{1}{N_S} \Four{p_{t_0}} \sum_{k=1}^{N_S} \bigg\{\left(\Four{ \sqrt{2\pi} p_0^\prime}\right)^k \nonumber\\
        &&{} \times \Four{ \sqrt{2\pi} p_\mu }( k\omega )\bigg\},\\
\int df \ p\left(f\right) \left| \Four{f}\right|^2 &=& \frac{1}{2\pi} \frac{1}{N_S} \Bigg( 1 + \sum_{k=1}^{N_S-1} \bigg[2\left(\frac{N_S-k}{N_S}\right)\\
        &&{}\hspace*{-1.5cm} \times \Re\left\{ \left( \Four{\sqrt{2\pi} p_0^\prime} \right)^k \Four{ \sqrt{2\pi} p_\mu }( k\omega ) \right\} \bigg]\Bigg).\nonumber
\end{eqnarray}

In our specific model, Eq. \ref{eq:iei} implies 
$p_0^\prime(\tau) \sim \mathcal{N}(0, \sigma_{\rm{jit}})$ and
$p_\mu(\mu) \sim \mathcal{N}( \mu_0, \sigma_\mu)$.  If we
draw the initial temporal offsets from $p_{t_0}(t_0) \sim  \mathrm{unif}(-\mu_0/2,\mu_0/2)$, this yields
\begin{widetext}
\begin{eqnarray}
\int df\ p\left(f\right) \Four{f}
   &=& \frac{1}{\sqrt{2\pi}}\frac{1}{N_S}\ \sinc\left( \frac{\mu_0\omega}{2} \right) \sum_{k=1}^{N_S}  e^{ -(k\sigma_{\rm{jit}}^2+k^2\sigma_\mu^2)\omega^2/2} e^{-ik\mu_0\omega},\\ \nonumber
\end{eqnarray}
\begin{eqnarray}
\int df\ p\left(f\right) \left| \Four{f}\right|^2
    &=& \frac{1}{2\pi} \frac{1}{N_S} \left( 1 + 
          \sum_{k=1}^{N_S-1} 2\left(\frac{N_S-k}{N_S}\right) e^{ -(k\sigma_{\rm{jit}}^2+k^2\sigma_\mu^2)\omega^2/2} \cos\left(k\mu_0\omega\right) \right).
\end{eqnarray}

Note,
\begin{eqnarray}
\left|\int df\ p\left(f\right) \Four{f}\right|^2
   &=& \frac{1}{N_S^2}\, \sinc^2\left( \frac{\mu_0\omega}{2} \right) \sum_{k=1}^{N_S}\bigg( 
       e^{ -(k\sigma_{\rm{jit}}^2 +k^2\sigma_\mu^2)\omega^2 }
            \nonumber+ 2 \sum_{\ell=k+1}^{N_S} \cos\left((\ell-k)\mu_0\omega\right)e^{ -((k+\ell)\sigma_{\rm{jit}}^2 +(k^2+\ell^2)\sigma_\mu^2)\omega^2/2 } \bigg).
\end{eqnarray}
Putting this together yields
\begin{align}
\label{eqn:finalexpectation}
   &\Expect{f}{\left|\Four{f_T}\right|^2} 
     = \frac{1}{2\pi}  \frac{N_C}{N_S} 
   \left( 1 + \sum_{k=1}^{N_S-1} 2 \left(\frac{N_S-k}{N_S}\right)\, \cos\left(k\mu_0\omega\right) e^{ -(k\sigma_{\rm{jit}}^2 + k^2\sigma_\mu^2)\omega^2/2 }\right) \nonumber \\
&+ \frac{1}{4\pi}\frac{N_C(N_C-1)}{N_S^2}\, \sinc^2\left( \frac{\mu_0\omega}{2} \right) \sum_{k=1}^{N_S}\Bigg(
       e^{ -(k\sigma_{\rm{jit}}^2 +k^2\sigma_\mu^2)\omega^2 } + 2 \sum_{\ell=k+1}^{N_S} \cos\left((\ell-k)\mu_0\omega\right)e^{ -((k+\ell)\sigma_{\rm{jit}}^2 +(k^2+\ell^2)\sigma_\mu^2)\omega^2/2 } \Bigg). 
\end{align}
\end{widetext}

\begin{figure}
\includegraphics{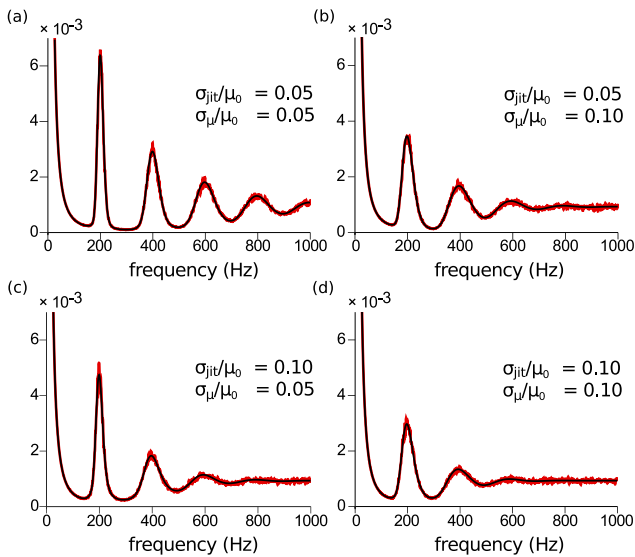}
\caption{Comparisons of analytically derived energy spectral density, Eq. \ref{eqn:finalexpectation} (black line), against
numerically computed energy spectral density (red line). Note the excellent match between these results. 
Energy spectra are normalized over the range $10$ to $1000$ Hz. Each plot shown
reflects the fixed parameters $\mu_{0}=5 \; \mbox{ms}$, $N_{C}=500$ cells,
and $N_{S}=500$ spikes. Numerical results are averaged over 500 simulations.}
\label{fig:dirac_delta}
\end{figure}

Eq.~\ref{eqn:finalexpectation} is the final expression for the energy spectral density of the train of delta functions
whose event times are specified by Eq.~\ref{eq:iei}.
Note that this result depends on five model parameters: 
$N_S$, the number of spikes per cell; $N_C$, the number
of cells in the population; $\mu_0$ and $\sigma_\mu$, which together determine $\mu$ for each cell;
and $\sigma_{\rm{jit}}$, which introduces variability from event to event (i.e., ``jitter''). 
Fig.~\ref{fig:dirac_delta} shows an excellent match between this analytical result 
and numerical simulations of the event train for several parameter combinations.

\begin{figure}
\includegraphics{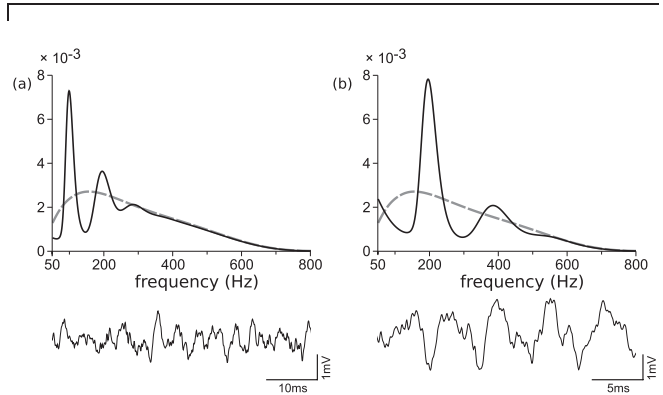}
\caption{Energy spectra (top) and example time-domain waveforms (bottom) for
action potential (AP)-convolved model LFP signals. Spike time variability parameters
were set to $\sigma_{\mu}/\mu_{0}=0.1$ and $\sigma_{\rm{jit}}/\mu_{0}=0.1$,
with (a) $\mu_{0}=10 \; \mbox{ms}$ and (b) $\mu_{0}=5 \; \mbox{ms}$. Normalized LFP energy
spectra described by Eq. \ref{eqn:finalexpectation} (solid; black) were compared against 
the energy spectrum of the AP waveform (dashed; gray), which is also the energy spectrum of
an AP-convolved Poisson process (white noise). Energy spectra were normalized to the range of
$50$ to $1000$ Hz. Note stronger emergent rhythms with smaller $\mu_{0}$ (higher frequency).}
\label{fig:ap}
\end{figure}

To make comparisons with experimental LFP recordings, the event train must be convolved with
a realistic voltage waveform, resulting in the model LFP spectrum being the product of the event train
spectrum and the waveform spectrum. Fig.~\ref{fig:ap} shows the results of convolving with a typical action potential (AP)
waveform, for $\mu_0$ corresponding to 100 Hz and 200 Hz. The model LFPs feature strong peaks in their spectra
at both frequencies, and voltage traces from numerical simulations show a clear rhythm in the 200 Hz
signal, demonstrating our primary point: completely asynchronous spiking may produce narrowband LFP rhythms
when neural activity is quasi-periodic. The 100 Hz oscillation is not as obvious in the time domain 
because a large proportion of energy is concentrated as high-frequency noise at 300+ Hz. 

\begin{figure}
\includegraphics{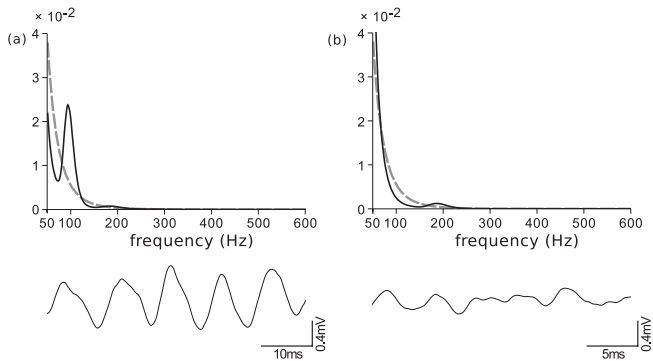}
\caption{Example energy spectra (top) and time-domain waveforms (bottom) for
postsynaptic potential (PSP)-convolved model LFP signals. Parameters and normalization
were set as in Fig. \ref{fig:ap}, with spectra of Eq. \ref{eqn:finalexpectation} (solid; black) compared against
the spectrum of the PSP waveform (dashed; gray), which is also the energy spectrum of a PSP-convolved 
Poisson process (white noise). Note stronger emergent
rhythms with larger $\mu_{0}$ (lower frequency).}
\label{fig:psp}
\end{figure}

Fig.~\ref{fig:psp} shows results from convolving the event train with a typical PSP waveform. Note how the energy of the PSP waveform is concentrated 
at much lower frequency than that of the AP waveform (gray dashed lines in Figs.~\ref{fig:ap} and \ref{fig:psp}), resulting in the 200 Hz PSP signal being severely attenuated compared to the 100 Hz PSP signal. This provides a simple explanation for the conventional wisdom that PSPs tend to dominate the LFP at low frequency, while APs tend to dominate at higher frequency \cite{schomburg2012spiking}. 

\begin{figure}
\includegraphics{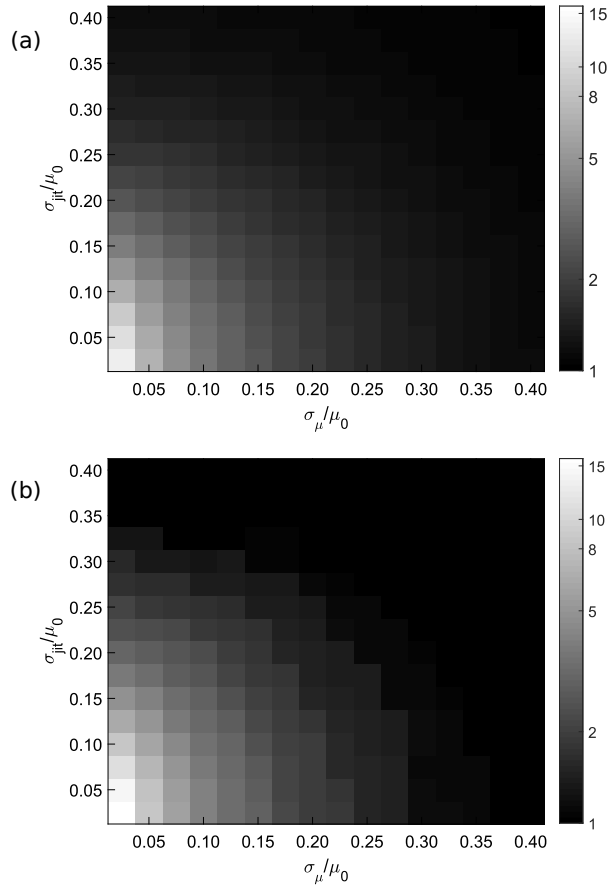}
\caption{Signal-to-noise ratio as function of spike time variability. (a) AP-convolved
signal with $\mu_{0}=5 \; \mbox{ms}$. (b) PSP-convolved signal with $\mu_{0}=10 \; \mbox{ms}$.
Note signal degradation is greater with increasing $\sigma_{\mu}/\mu_{0}$
against $\sigma_{\rm{jit}}/\mu_{0}$. Emergent rhythms depreciate beyond
noticable detection when spike time variability ratios are each $\gtrsim 0.25$.}
\label{fig:colorplot}
\end{figure}

To characterize the strength of rhythms emerging from asynchronous
neural activity, in Fig.~\ref{fig:colorplot} we plot the
signal-to-noise ratio (SNR) as a function of $\sigma_{\mu}$ and
$\sigma_{\rm{jit}}$ for 200~Hz AP-convolved LFPs and 100 Hz
PSP-convolved LFPs. We define the SNR as the ratio of the LFP energy
spectral density to the waveform energy spectral density at
$f=\frac{1}{\mu_0}$. The waveform energy spectral density is
considered the noise spectrum since it is what would result from a
Poisson event train (white noise)
convolved with the waveform. Note how $\sigma_{\mu}$ (population heterogeneity) and
$\sigma_{\rm{jit}}$ (IEI heterogeneity) do not have the same effect on SNR---increasing
$\sigma_{\mu}$ degrades SNR more quickly than increasing
$\sigma_{\rm{jit}}$, as a result of its being attached to a factor
of $k^2$ rather than $k$ in Eq.~\ref{eqn:finalexpectation}. Our model
therefore predicts that heterogeneity in mean firing frequency across
a neural population will degrade asynchronous rhythms more than an
equivalent degree of spike-time jitter. The results in
Fig.~\ref{fig:colorplot} also suggest bounds on these two sources of
spike-time variability for facilitating the emergence of LFP rhythms
from asynchronous neural activity. For both AP events at 200 Hz and
PSP events at 100 Hz, $\sigma_{\mu}$ and $\sigma_{\rm{jit}}$ can
each reach as high as about 25\% of $\mu_0$ before the primary
spectral peak is washed out by noise.

Our model therefore makes three main predictions. First, completely asynchronous and independent neural activity may produce robust, narrowband LFP oscillations, so long as individual neural activity is quasi-periodic. (Note that quasi-periodicity is essential---independent Poisson processes, for example, result in a flat power spectrum \cite{poisson}, but in many cases do not accurately describe neural activity \cite{reyes2003synchrony,cateau2006relation}.) Previous computational work supports this hypothesis, suggesting that pathological ``high-frequency oscillations'' associated with epileptic seizures may be generated by a completely asynchronous, uncoupled network of hippocampal pyramidal cells receiving intense synaptic input \cite{fink2015network}. Second, rhythms generated by asynchronous activity are degraded more by heterogeneity in intrinsic neuronal frequency than by neuronal jitter. And third, we have derived bounds on these two sources of heterogeneity for experimentally detecting oscillations from asynchronous neural activity. Our model additionally provides a simple mathematical explanation for why PSP waveforms tend to dominate the LFP at low frequency, while AP waveforms dominate at high frequency. These results should spur future experimental studies which investigate the possibility of neural oscillations emerging from asynchronous neural activity, especially under pathological conditions such as epileptic seizures.

\subsection{}
\subsubsection{}

\begin{acknowledgments}
This work was supported by NIH Grant Nos. R01-NS094399, K08-NS069783, UL1-TR000433, and K01-ES026839. We also acknowledge funding from the Doris Duke Charitable Foundation Career Development Award, NSF Grant No. 1003992, and the Ohio Wesleyan Summer Science Research Program. Special thanks to Bob Harmon for suggesting the helpful optics analogy.
\end{acknowledgments}

\bibliography{asynch_narrowband_oscillations_arxiv}

\end{document}